\documentclass[10pt,prb,twocolumn,showpacs]{revtex4-1}
\usepackage{graphicx}
\usepackage{dcolumn}
\usepackage{amsmath}
\usepackage{bm}
\usepackage{tikz}
\usepackage[normalem]{ulem}

\usepackage[colorinlistoftodos,prependcaption,textsize=tiny]{todonotes}

\begin{document}

\title{Electric field induced structural changes of water confined between two graphene layers}
\author{Sobrino Fern\'andez Mario, M. Neek-Amal\email{neekamal@srttu.edu} and F. M. Peeters}
 \affiliation{Departement Fysica,
Universiteit Antwerpen, Groenenborgerlaan 171, B-2020 Antwerpen,
Belgium}
\date{\today}
\begin{abstract}

An external electric field changes the physical properties of polar-liquids due to the reorientation of their permanent dipoles. For example it should affect significantly the physical properties of water confined in a nanochannel. The latter effect is profoundly enhanced, if the field is applied along the nanochannel. Using molecular dynamics simulations, we predict that an in-plane electric field applied parallel to the channel polarizes water molecules which are confined between two graphene layers, resulting in distinct-ferroelectricity and electrical hysteresis. We found that electric fields alter the in-plane order of the hydrogen bonds: reversing the electric field does not restore the system to the non-polar initial state, instead a residual dipole moment remains in the system. Our study provides insights into the ferroelectric state of water when confined in nanochannels and shows how this can be tuned by an electric field.

\end{abstract}
\pacs{64.70.Nd}

\maketitle


\section{Introduction}~

Whether or not there exists a phase of ice with a net molecular dipole moment is an
old question which is closely related to the debate on the existence of ferroelectricity in
ice. It remains highly controversial because of the absence of both theoretical and experimental study on the microscopic
structure of water in the presence of an electric field. There is experimental evidence that at least partial
ferroelectric alignment can be induced in normal ice, either by
interaction with a substrate~\cite{0,1} or by doping with
impurities~\cite{03}. Su~\emph{et al}.~\cite{1} found ultra-thin
hexagonal films of water covering a substrate (1-10 monolayer thick) which had a
net dipole moment. Ferroelectricity is expected to appear in the
ordered phase of ice where the hydrogen bonds (H-bonds) are aligned uniformly
~\cite{04}. Importantly, when reducing the degrees of freedom of 
the H-bonds the system becomes more ordered.

The ferroelectricity of monolayer ice was studied by several
groups which found conflicting results, e.g. hexagonal and 
flat/rippled rhombic phases were found to be ferroelectric by
Zhao~\emph{et al}~\cite{Zhao}. This ordered phase has a net dipole. However many other studies reported a
 disordered phase with zero net dipole moment~\cite{arxiv2015,Kaneko,PhysRevE.72.051503}.

An external electric field, should in principle, reorient the local dipoles of water. However because of the random
distribution of the dipoles in ordinary water, the strength of the electric field should be sufficiently strong ($\geq 0.1$ V/\AA) to disrupt the network of molecules in liquid water~\cite{dis} and to change the equilibrium freezing point of water.
A higher electric field of about $>0.36$ V/\AA~ causes proton flow in hexagonal ice~\cite{icemove}.
Depending on the value of the electric field, the external pressure may cause melting or freezing of water~\cite{acta}.
Recently, it was proposed experimentally that confined water exists as a quasi-two-dimensional
layer with different properties than those of bulk water~\cite{apl2015,nat2015}, though this observation was challenged later~\cite{NatureComment}. Graphene, the two-dimensional
allotrope of carbon~\cite{Novoselov}, was used in a recent experiment to  confine water~\cite{nat2015} into monolayer, bilayer
and trilayer. In our previous study we showed that a flat monolayer of ice exists between two graphene layers~\cite{PhysRevB.92.245428}.
Here we address the effects of an electric field on a monolayer of ice which is confined between two graphene layers (separated by a few Angstr\"om). The microscopic details of the orientation of the H-bonds is interesting and provides insights about the response of ice to an electric field in other dimensions.

\begin{figure}[b]
\includegraphics[width=0.4\textwidth]{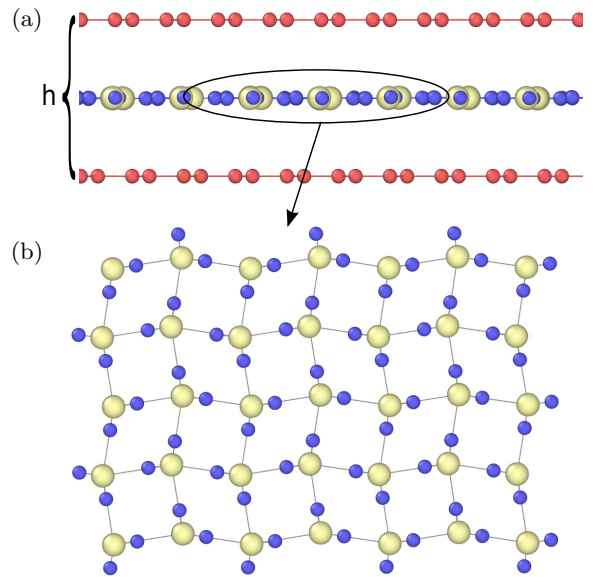}
\begin{tikzpicture}[overlay]
\draw (-7.5,7.6) node {(a)};
\draw (-7.5,4.5) node {(b)};
\end{tikzpicture}
\caption{(color online) The top (a) and side (b) view of a relaxed monolayer of ice between two graphene layers.
\label{fig1} 
}
\end{figure}

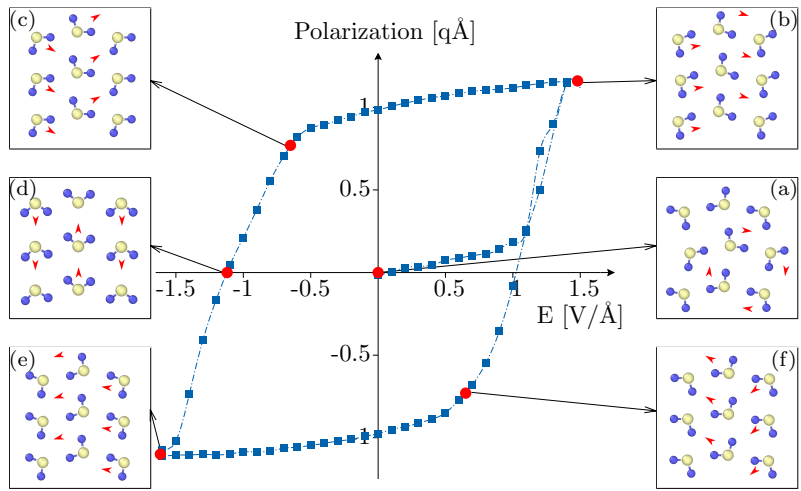
\begin{figure*}
\includegraphics[width=0.6\textwidth]{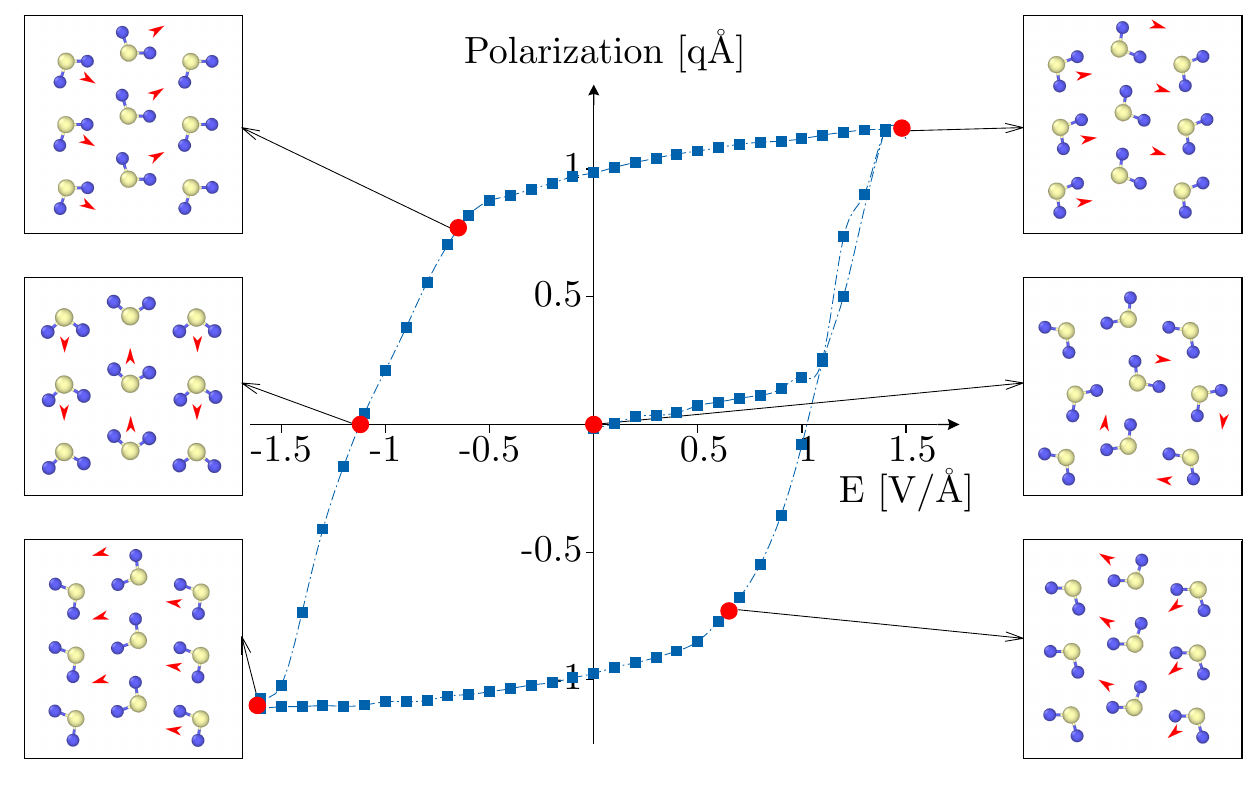}
\begin{tikzpicture}[overlay]
\draw (-0.36,4.23) node {\footnotesize{(a)}};
\draw (-0.36,6.48) node {\footnotesize{(b)}};
\draw (-10.48,6.48) node {\footnotesize{(c)}};
\draw (-10.48,4.23) node {\footnotesize{(d)}};
\draw (-10.48,1.95) node {\footnotesize{(e)}};
\draw (-0.36,1.95) node {\footnotesize{(f)}};
\end{tikzpicture}
\caption{\label{Fig:H65A} (color online) Hysteresis curve of the net polarisation of water confined between two graphene layers which are separated by $h$=6.5\AA~. The insets show the structure of monolayer ice for six specific cases. The electric field was varied with a rate of $12.5$ GHz 
\newline
}
\end{figure*}

Using the reactive force-field (ReaxFF~\cite{reax}), we
confirmed that the minimum energy configuration of a flat monolayer
ice confined between two graphene layers separated by 6.5\,\AA~ is a
non-polar ordered phase consisting of a combination of rhombic
and tilted square lattice structure. This structure
satisfies the ice rule while it is flat, ordered and non-polar.
We study the response of such a monolayer ice to an external in-plane 
electric field. We found that an electric field of about 1-2\,V/\AA~ rotates the H-bonds,
and reorients the dipoles changing the structure of monolayer ice. By
reversing the electric field direction we found irreversibility in
both the structure and the polarization of the system.


\section{Computational method}

In our MD simulations we used  the reactive force-field ReaxFF~\cite{reax} which is a
general bond-order dependent potential that provides an accurate description of chemical reactions. The connectivity in
the entire system is recalculated in every iteration and non-bonded interactions (e.g., van der Waals) are calculated between
all atom pairs~\cite{reax,Newsome,PRB2013}.
Numerical simulations are carried out using the LAMMPS code~\cite{Kresse} which includes the ReaxFF method~\cite{Zybin}.
The ReaxFF potential allows H-bond extension/contraction in water as well as angle bending and it allows charge
relaxation over each atom. This is in contrast to the traditional force fields for water, e.g. SPC and TIP4P~\cite{tip2005} (a rigid planar
four-site interaction potential for water) that keep the water molecules rigid during MD simulations.
Notice that ReaxFF takes into account charge relaxation for all atoms in the system including carbon atoms.
The time step used in the simulations is $2\,\times 10 ^{-7}$ ns and each simulation consists of $10^6$ time steps ($0.2$ ns).

%



\section{Minimum energy configurations}~~

The computational unit cell is a square with dimension $283\times163\AA^2$ that contains 34848 carbon atoms and filled by
17100  water molecules. We start from a dense square structure of O atoms with a lattice constant of 2.8\AA~ and a random distribution of the H-bonds. The monolayer of water is confined between two
graphene layers and interaction with the C-atoms is fully taken into account. We performed an annealing MD simulation using NPT starting at 400\,K and ending at 0\,K, in order to find the true simulation box size
and O-O distances. The total energy is minimized using the iterative
conjugate gradient (CG) scheme. The graphene layers are rigid and
separated by a fixed distance (h) having AB-stacking. More details
about our method can be found in Ref. \cite{PhysRevB.92.245428}.

We found a flat rhombic-square lattice structure for monolayer ice confined between graphene layers that are separated by $h$=6.5\AA,~ see Fig.~\ref{fig1}.
~The top view of the minimum energy configuration is shown in
Fig.~\ref{fig1} which is a flat monolayer of ice with successive
arrangements of rhombic building blocks without a net dipole moment.
The orientation of the H-bonds leads us to conclude that we found an ordered phase without a net dipole moment.
 The radial distribution function (RDF) shows that  the O-O distance in flat ice is about a=2.84$\pm$0.01\AA~which is
in good agreement with the experimental value of a=2.81$\pm$0.02\AA \cite{nat2015}.

\begin{figure}
\includegraphics[scale=0.57]{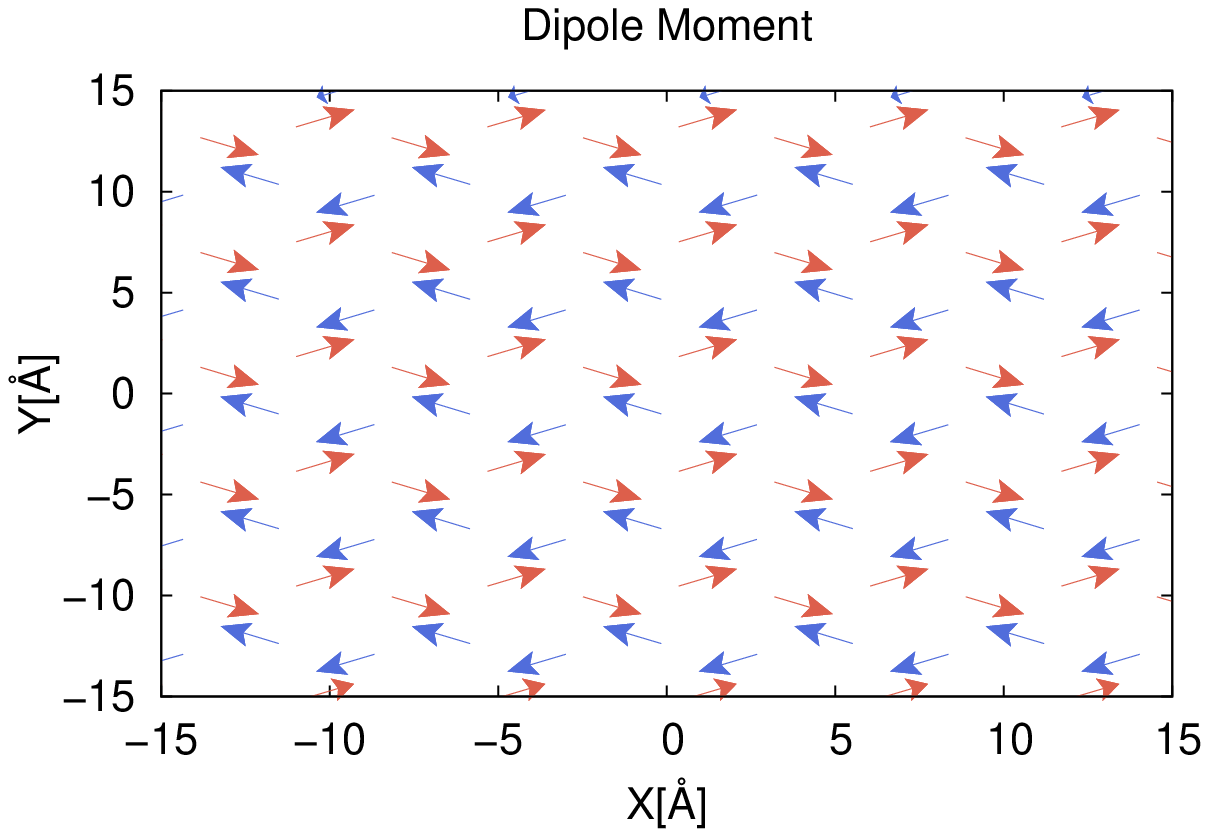}
\includegraphics[scale=0.57]{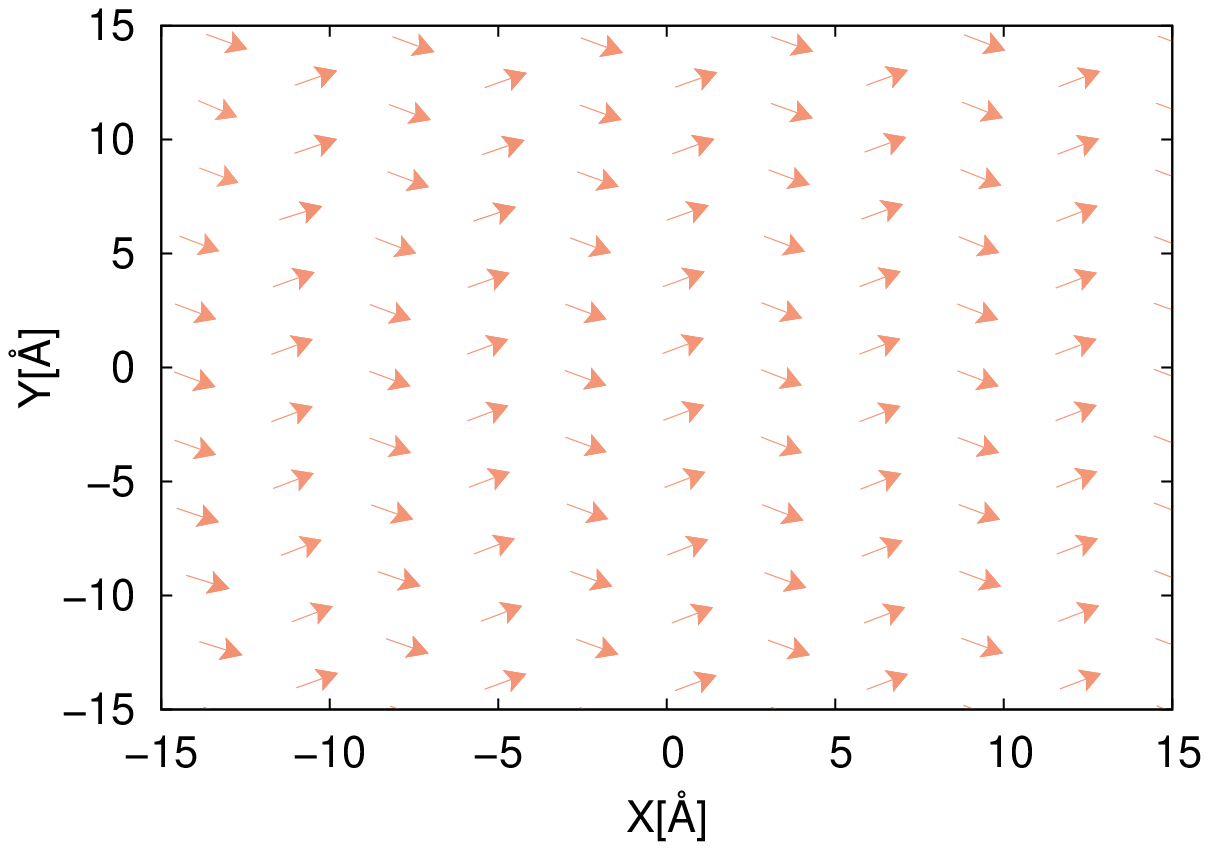}
\caption{(color online) Arrows indicate the local dipoles in the monolayer of water confined between two graphene layers for (a) a non-polarised state, and (b) at the saturation polarisation. 
\label{Fig:Dipoles}
}
\end{figure}

\section{Electrical hysteresis loop}~~

In order to investigate the ferroelectric properties of a monolayer ice
confined between two graphene sheets, we measure the total dipole moment
by summing over all local dipoles ($\vec{P}_{total}=\sum_{n} \vec{p}_n$) in the presence of an
external in-plane electric field of the order V/\AA~in the zig-zag direction.



By changing the electric field in the range [-1.5,1.5]V/\AA~ we found hysteretic behaviour which is explained as follows. For a tightly confined ice crystal ($h\in$[4.5-6.5]\AA)~ we find a smooth hysteresis curve, which for $h$=6.5\AA~is shown in Fig.~\ref{Fig:H65A}. In the insets of Fig.~\ref{Fig:H65A} the corresponding ice structures for six special states are depicted.  Inset (a) shows the minimum energy configuration of a monolayer of water (Fig. 1). By increasing the electric field the system becomes polarized and one naturally expects that the local dipoles orient along the direction of the external  field. However, such a complete alignment does not occur in monolayer ice because of the local dipole-dipole interaction as well as the strong (mostly Coulomb) interaction between the neighbours. 
The minimum energy is determined by a compitition between the external electric field and the local arrangement of the dipoles. Surprisingly all the O atoms remain in the square-rhombic structure and only the H-bonds are rotated in the presence of the electric field.  In the inset (b) the water configuration is shown near the saturation point $P_s$. Two columns are visible for which the perpendicular dipole orientation is opposite, reminiscent of H-bonds between neighbouring water molecules. When decreasing the electric field after reaching the saturation point, the system retains a large fraction of polarization until a critical field is reached ($+P_r$).
Here, the system reacts similar to a 2D Ising model: it relaxes through avalanche-like dynamics shown in the insets (c-d-e).
We conclude that due to the presence of H-bonds, the larger electric field can only rotate the H-bonds and there is no significant change in the location of the oxygen atoms.
By changing the electric field as seen from the insets there are two distinguishable lines of water molecules with a different orientation of the local dipoles. The local dipoles in two adjacent columns in the case $\pm E_c$ are in opposite directions which results in a zero net dipole moment. We can conclude that electric fields $<2$V/A~are weak enough to change the position of O atoms. Changing the position of O atoms demands higher electric field which deforms the ice structure. 



\section{Effect of the rate of electric field}

\begin{figure}[b]
\includegraphics[scale=0.7]{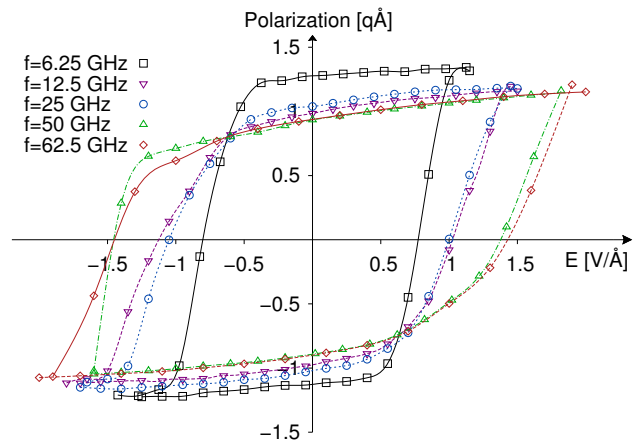}
\caption{\label{Fig:HystAllfreq} (color online) Hysteresis loop of the polarization of an ice monolayer confined between two graphene sheets for different frequencies of the applied electric field. }
\end{figure}

It is well known that hysteresis effects are related to the starting configuration and gives a time dependent output. Therefore we expect that by changing the rate (frequency) at which the electric field is varied we obtain different hysteresis loops. The electric field is varied linear in time between $-2$ V/\AA~and $+2$ V/\AA~ with different rates f. 
In figure~\ref{Fig:HystAllfreq} we performed several additional simulations using different values of the frequency of the external field. For instance the frequency f=12.5 GHz corresponds to the situation which was shown in more detail in Fig. \ref{Fig:H65A}. It is important to note that changing the frequency of the external field does not effect the morphology of the structures, only the rate at which the changes take place. The system reaches its highest saturation and remnant polarization for the slowest considered input rate of $6.25$ GHz. By increasing the frequency these values gradually decreases, and the critical field for which the system switches its polarization is increased.
There are three parameters that characterises each loop: the field $E_c$ at zero dipole moment,  the remnant polarization $P_r$ and the saturation polarization $P_s$ which are listed in Table~\ref{Table:HystV}. From this table we see that there are different frequency ranges for which the polarization process are rather similar. In general we can conclude that the saturation polarization for higher frequencies lowers, and that the driving electric field must be increased to a larger value before the system switches between polarized states.
In Fig. \ref{Fig:Dipoles}(a)  the local dipoles are shown for the unpolarized state (see inset (a) in Fig. \ref{Fig:H65A} for the corresponding orientation of the water molecules). The local dipole moment of each molecule is compensated by molecules at the opposite vertex of the lattice. When applying the external field this symmetric arrangement is no longer preserved, and we find a new structure as shown in Fig. \ref{Fig:Dipoles}(b). Here we find columns of similar orientation in the lateral direction of the electric field, with the total dipole moment averaging out in the direction of the external field. To preserve the H-bonds in the rhombic lattice, the individual dipolar molecules never fully align themselves with the external field. When increasing the strength of the external field to $\pm 3$ V/\AA,~there is a transition to a triangular lattice where all the local dipoles point in the direction of the field.
In summary by increasing the frequency the width of the loops is enlarged  and we obtain larger (smaller) $E_c$ ($P_r$).

\begin{table}
\begin{tabular}{|c|c|c|c|c|}
\hline
  f [GHz]    & $P_{s}[q\AA] $&    $P_r [q\AA]$ &   $E_c[V/\AA]$  \\
\hline
$6.25$  &   $1.28 \pm 0.02$   &   $1.20 \pm 0.02$   &   $0.79 \pm 0.02$  \\
\hline
$12.5$  &  $1.15 \pm 0.03$   &  $0.98 \pm 0.02$  &   $1.05 \pm 0.06$  \\
$25$      &$1.17 \pm 0.04$   & $1.02 \pm 0.02$   & $1.03 \pm 0.06$  \\
\hline
$50$      &  $1.10 \pm 0.02$   &  $0.91 \pm 0.05$   &   $1.43 \pm 0.02$  \\
$62.5$   & $1.12 \pm 0.02$ &   $0.92 \pm 0.05$  &   $1.47 \pm 0.02$  \\
\hline
\end{tabular}\par
\caption {Characteristic hysteresis parameters for different
frequencies: Coercive field ($E_c$), remnant
polarization ($P_r$), and saturation polarization ($P_s$).}
\label{Table:HystV}
\end{table}

\begin{figure}[t]
\includegraphics[scale=0.57]{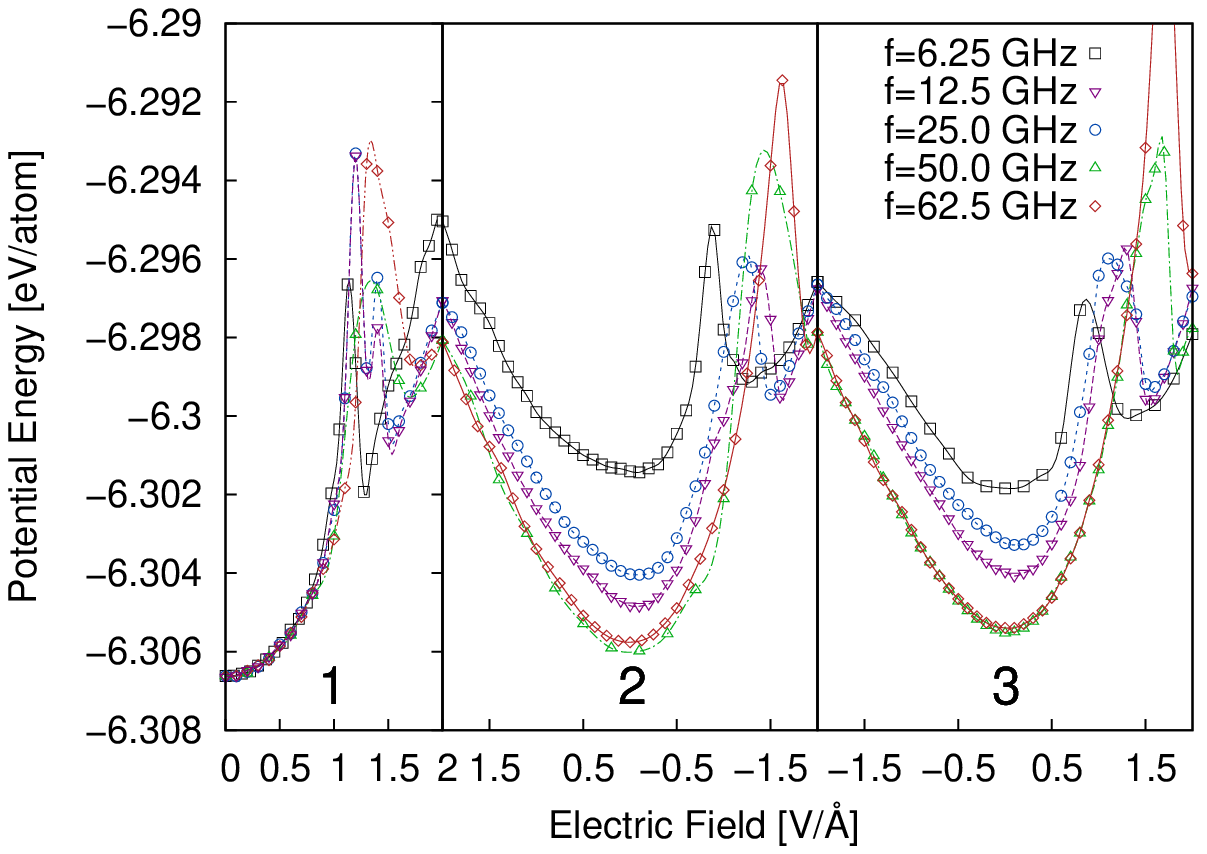}
\includegraphics[scale=0.57]{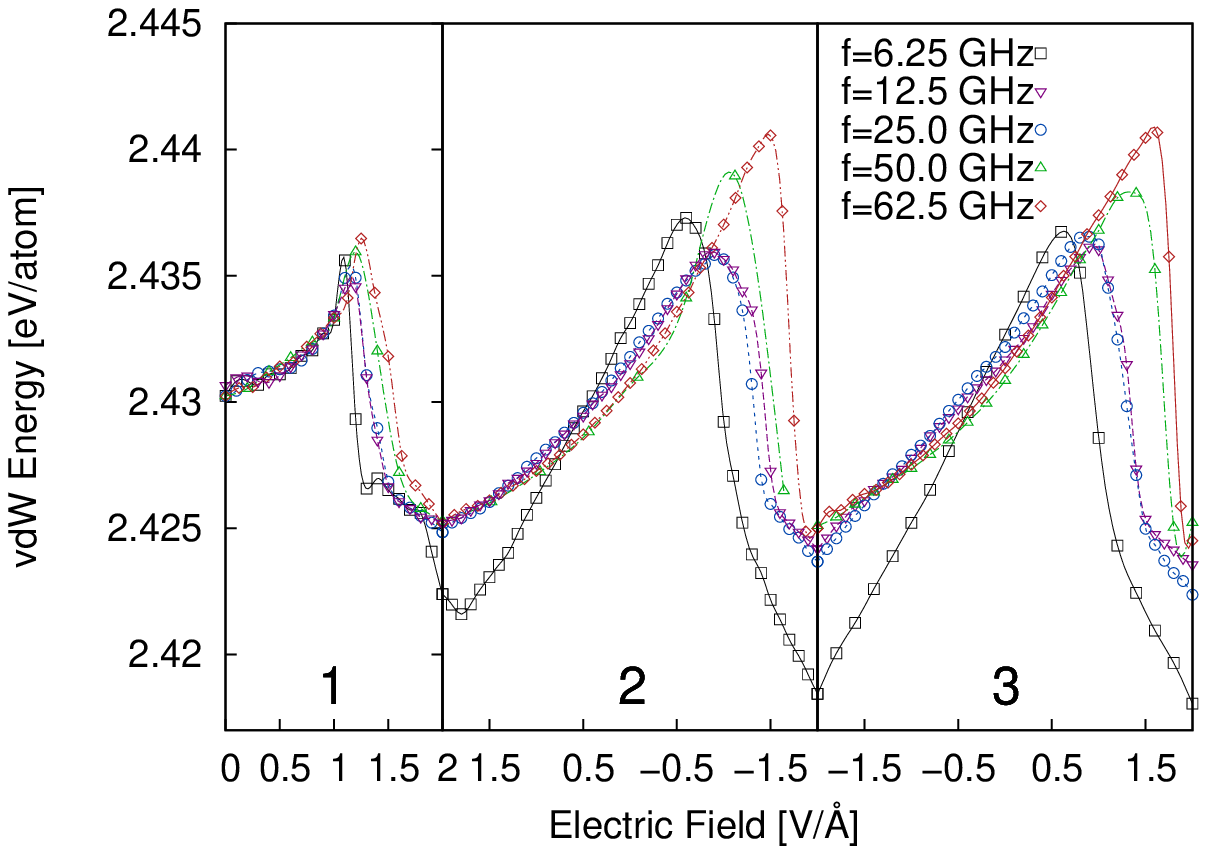}
\caption{\label{Fig:DifHzE} (color online) Potential (a) and van der Waals (b) energy as function of applied electric field for different frequencies of the electric field. Three domains in (a) and (b) refer to three different steps: (1)  increasing the dipole starting from a zero field value, (2) lowering the external field up to $\simeq$-1.5V/\AA,~and (3) increasing the external electric field up to $\simeq$+1.5V/\AA.}
\end{figure}

For all studied $h$ values, confined ice has the same energy threshold value for which it responds to the external field and passes from one metastable state to another. When increasing the rate of the external field, this barrier increases.
The potential energy of the system is shown in Fig.~\ref{Fig:DifHzE} (a). We find that the energy oscillates with the same frequency as the electric field and that for all given rates, a large jump in energy is visible around the field for which the system switches its polarization. For high frequency fields, the energy of the remnant state is closer to the initial configuration.
The van der Waals energy is shown in Fig. \ref{Fig:DifHzE}(b). Here, the tension on the dipoles created by switching the polarity of the field is clearly visible. The vdW energy reaches a minimum when we reach a maximum alignment of the local dipoles, which is at the saturation point (end of domains (1,2,3) in Fig. \ref{Fig:DifHzE} (b)). Comparing this result with structures found in \ref{Fig:H65A} we see that here the H-bonds between neighboring atoms is the strongest. When reaching the critical field, the H-bonds are weakened as the individual columns rotate towards a perpendicular alignment with the electric field. The eventual cascade towards a switched polarization is prolonged for higher frequencies, and the vdW energy reaches larger values for these systems.


\section{Discussions}

Ferroelectric materials are used in many areas of science and technology which are a special
class of piezoelectrics showing a large piezoelectric response. 
Ferroelectricity of confined water is largely unknown because of
the lacking of experiments. The control of dipoles in water,
its description and understanding of electromechanical (piezoelectric)
and ferroelectric hysteresis helps to understand the fundamental
properties of confined water and its fluidity \cite{Mehdiarxiv}.  Recently there have been
several experimental attempts to produce confined monolayer ice and
to study confined water in a hydrophobic channel~\cite{nat2015}. Here theoretically we
obtained the minimum energy configuration for confined ice and
its irreversibility response to an in-plane electric field. We found
that the minimum energy configuration is not a square
lattice structure. An in-plane electric field rotates the H-bonds and makes
monolayer ice polarized (but the oxygen atoms remains almost fixed)
without losing the flatness of monolayer ice. However, by
reversing the field the system reaches a higher energy state.
The higher energy configuration mostly consists of two columns of dipoles whose moment can cancel each other when the electric field is set to $\pm E_c$.
Therefore, monolayer ice oscillates between two ordered phases with opposite polarity (Fig.~\ref{Fig:H65A} (b-e))
 when an alternating electric field is applied. Independent of the applied electric field
all structures remain almost flat. This electromechanical coupling
might be responsible for some unusual phenomena in confined ice.

\section{Conclusions}

In summary using reactive molecular dynamics, we studied systematically the response of monolayer ice to an external in-plane electric field.
The structural change, the vdW and the H-bonding energy were calculated for such a monolayer of ice. We discovered electrical hysteresis in this system, where the net dipole moment of the system changes irreversibly with the electric field. During this process the system remains almost flat, and the oxygen atoms maintain their original position. The microscopic details that we address here for the response of a two-dimensional ice lattice to an external electric field is helpful for the understanding of the response of other structures of ice  to an external electric field. We believe that this finding will motivate experimentalists to realize the proposed effect and reveal many unexplored issues of confined ice. Confined two-dimensional ice is a new material that exhibits hysteresis phenomena.

\textit{Acknowledgements} This work was supported by the Flemish Science
 Foundation (FWO-Vl) and the Methusalem Foundation of the Flemish Government.

\bibliography{refs}{}

\end{document}